\documentclass[letter]{aa}

\usepackage{graphicx}
\graphicspath{{./figures/}}
\usepackage{txfonts}
\usepackage{natbib}
\bibpunct{(}{)}{;}{a}{}{,}

\begin{document}

% abbreviations
\newcommand{\mC}{\mbox{\ensuremath{^{\rm 12}}C}}
\newcommand{\iC}{\mbox{\ensuremath{^{\rm 13}}C}}
\newcommand{\HII}{\mbox{H{\sc ii}}}
\newcommand{\CII}{\mbox{[C{\sc ii}]}}
\newcommand{\mCII}{\mbox{[\mC{\sc ii}]}}
\newcommand{\iCII}{\mbox{[\iC{\sc ii}]}}

\author{
  U.~U. Graf\inst{1}\and
  R. Simon\inst{1}\and
  J. Stutzki\inst{1}\and
  S.~W.~J. Colgan\inst{2}\and
  X. Guan\inst{1}\and
  R. G\"usten\inst{3}\and \\
  P. Hartogh\inst{4}\and
  C.~E. Honingh\inst{1}\and
  H.-W. H\"ubers\inst{5,6}
}
\institute{
  I. Physikalisches Institut der Universit\"at zu K\"oln, Z\"ulpicher
  Stra{\ss}e 77, 50937 K\"oln / Germany; graf@ph1.uni-koeln.de
  \and
  NASA Ames Research Center, Moffett Field, CA, USA
  \and
  Max-Planck-Institut f\"ur Radioastronomie, Auf dem H\"ugel
  69, 53121 Bonn / Germany
  \and
  Max-Planck-Institut f\"ur Sonnensystemforschung,
  Max-Planck-Stra\ss e 2, 37191 Katlenburg-Lindau / Germany
  \and
  Deutsches Zentrum f\"ur Luft- und Raumfahrt,
  Institut f\"ur Planetenforschung,  
  Rutherfordstra\ss e 2, 12489 Berlin / Germany
  \and
  Institut f\"ur Optik und Atomare Physik, Technische Universit\"at Berlin, 
  Hardenbergstra\ss e 36, 10623 Berlin / Germany
}
\title{\mCII\ and \iCII\ 158 $\mu$m emission from NGC 2024:\\
Large column densities of ionized carbon}
\date{Received 31 January 2012 / Accepted 1 March 2012}
\abstract 
{%context 
We analyse the NGC 2024 \HII\ region and molecular cloud interface using
\mCII\ and \iCII\ observations.}
{%aim
We attempt to gain insight into the physical structure of the interface
layer between the molecular cloud and the \HII\ region.}
{%methods 
Observations of \mCII\ and \iCII\ emission at 158 $\mu$m with high
spatial and spectral resolution allow us to study the detailed structure of the 
ionization front and estimate the column densities and temperatures of the 
ionized carbon layer in the PDR.}
{%results 
The \mCII\ emission closely follows the distribution of the 8 $\mu$m continuum.
Across most of the source, the spectral lines have two velocity peaks similar
to lines of rare CO isotopes. The \iCII\ emission is detected near the edge-on
ionization front. It has only a single velocity component, which implies that
the \mCII\ line shape is caused by self-absorption.  An anomalous hyperfine
line-intensity ratio observed in \iCII\ cannot yet be explained.}
{%conclusions
Our analysis of the two isotopes results in a total column density of 
N(H) $\approx$ 1.6$\times$10$^{23}$ cm$^{-2}$ in the gas emitting the \CII\
line. A large fraction of this gas has to be at a temperature of several
hundred K. The self-absorption is caused by a cooler (T$\le$100 K) foreground
component containing a column density of N(H) $\approx$ 10$^{22}$ cm$^{-2}$.}


\titlerunning{\mCII\ and \iCII\ 158 $\mu$m emission from NGC 2024}

\maketitle

%
%------Start of the normal Text ---------
%
% Introduction
\section{Introduction}
NGC 2024 is a well-studied star forming region at a distance of 415 pc
\citep{1982AJ.....87.1213A} in the Orion B complex. The source consists of a
dense ($n\approx 10^6$ cm$^{-3}$, \citet{1991A&A...246..570S}), narrow
($\approx$1$^\prime$), north-south extended molecular cloud with an embedded
\HII\ region
\citep{1982A&AS...48..345K,1986ApJ...307..302C,1989ApJ...342..883B}, which
extends several arcminutes beyond the molecular ridge in the east-west
direction. The molecular material in front of the \HII\ region is seen as a
prominent dust lane in the optical. Both OH \citep{1989ApJ...342..883B} and
H$_2$CO \citep{1986ApJ...307..302C} absorption line measurements relative to
the radio continuum indicate that this material is at a radial velocity of
$v_{\rm LSR}\approx9$ km/s. The bulk of the molecular gas behind the \HII\
region is found at $v_{\rm LSR}\approx11$ km/s from observations of e.g.
optically thin CO
\citep{1992A&A...256..640M,1993ApJ...405..249G,2009A&A...496..731E,2010MNRAS.401..204B}
or HCO$^+$ \citep{1989MNRAS.241..231R} lines. In the south of the \HII\ region,
a sharp ionization front delineates the boundary between the ionized and the
molecular gas (e.g. MSX data in \citet{2008AJ....136.1947W}). This ionization
front is seen edge-on in the south of the \HII\ region, but probably also
extends north in a face-on geometry along both the foreground and the
background parts of the cloud.

The velocity dispersion in each of the two main parts of the cloud is small
$\Delta v_{\rm LSR}\approx2$ km/s, which leads to a conspicuous double-peaked
line shape in many molecular emission lines. Detailed modeling of these lines
\citep{1993ApJ...405..249G,2009A&A...496..731E} results in H$_2$ column
densities of a few times 10$^{22}$ cm$^{-2}$ for the foreground component and
1--2$\times$10$^{23}$ cm$^{-2}$ for the background component. Gas temperatures
are found to be around 70 K in the background and around 30 K in the
foreground, with a likely increase near the \HII\ region interface.

In this {\it Letter}, we present \CII\ maps of an area near the ionization
front. Our analysis concentrates on the physical properties implied by the
strong \iCII\ emission detected.

\section{Observations}
We observed the $^2P_{3/2}\to {^2P_{1/2}}$ fine structure transition of ionized
carbon (C$^+$) at 1900.5369 GHz (157.7 $\mu$m) \citep{1986ApJ...305L..89C}
toward the NGC 2024 (Orion B) molecular cloud. The measurements were made on
Nov.\ 8, 2011 using the German REceiver for Astronomy at Terahertz frequencies
(GREAT)\footnote{ GREAT is a development by the MPI f\"ur Radioastronomie and
KOSMA/Universit\"at zu K\"oln, in cooperation with the MPI f\"ur
Sonnensystemforschung and the DLR Institut f\"ur Planetenforschung}.
\citep{2012A&A_specialv_Heyminck} on the Stratospheric Observatory for
Far-Infrared Astronomy (SOFIA) \citep{2012ApJ_Young}.

The receiver noise temperature during the observations was approximately 4000 K
(SSB). With a zenith opacity of 0.03--0.07, the system temperature was lower
than 4500 K (SSB). The beam size was 16\arcsec. The velocity resolution before
smoothing was 33 m/s.

We mapped a $192\arcsec\times144\arcsec$ area in total power on-the-fly mode.
Both the sampling distance along the scan direction and the scan spacing were
6$\arcsec$. The map was centered on ($\alpha,\delta$) = ($05^{\rm h}41^{\rm
m}45\fs20, -01\degr55\arcmin45\farcs0$) (J2000) using an off-position at
($05^{\rm h}41^{\rm m}36\fs30$, $-02\degr05\arcmin00\farcs0$). According to
\citet{1994ApJ...436..203J} and confirmed by a comparison measurement to a
second, far-away off-position, no \CII\ emission could be detected at this
position.

The measurements were calibrated using the procedure outlined in
\citet{2012A&A_specialv_Guan}. Instead of the main beam efficiency of 0.51
\citep{2012A&A_specialv_Heyminck}, we used a value of 0.58 to account for the
non-negligible coupling of the first side lobes to an extended source, as
derived from a model of the diffraction pattern. A linear baseline was
subtracted from each spectrum.

\section{Results}

\begin{figure}[btp]
    \begin{center}
      \includegraphics[width=9cm]{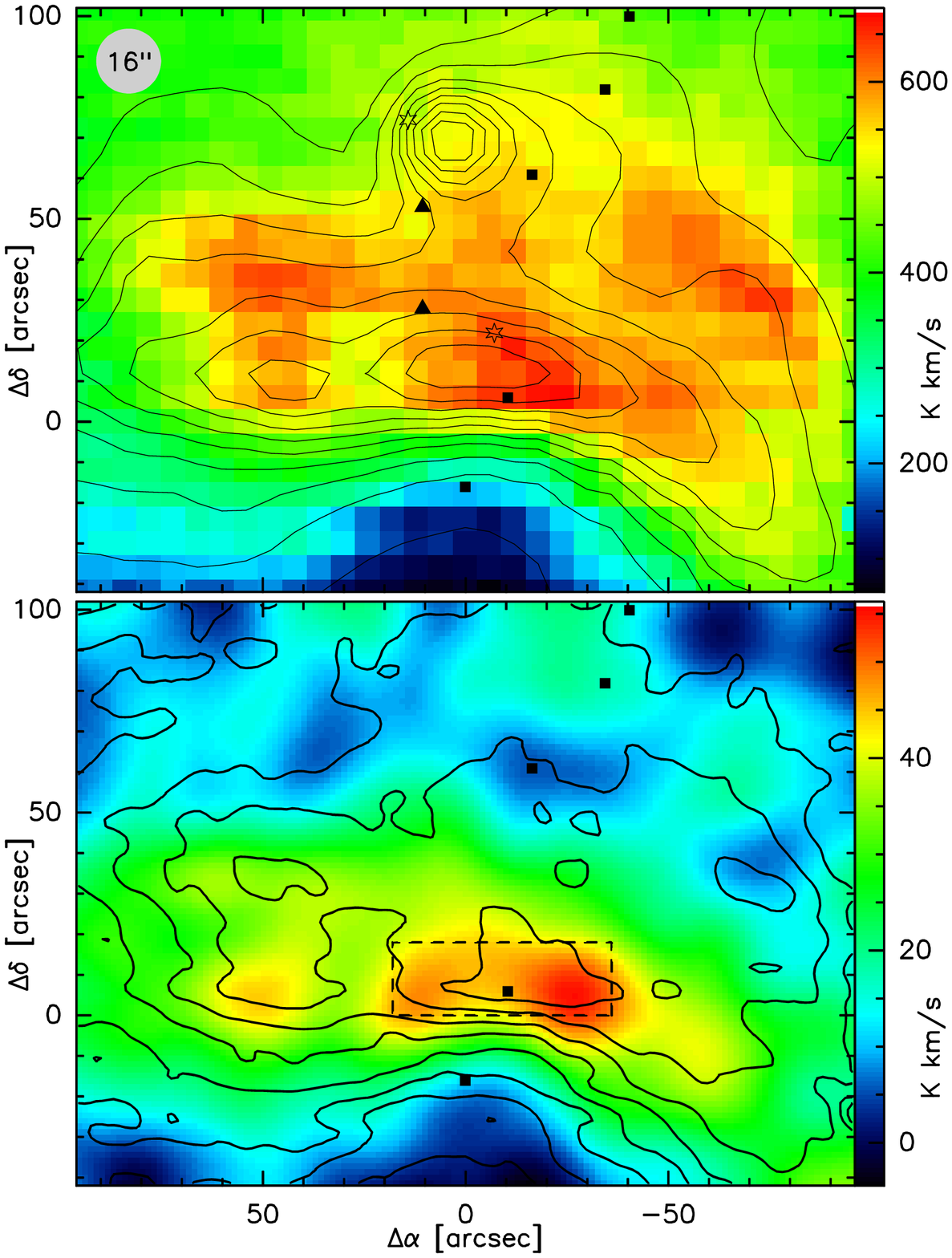}
     \end{center}
     \caption{
       \label{int_intens_map}
       {\bf Top:} Integrated (5--15 km/s) \CII\ intensity map of
       NGC 2024 (color code). Comparison data: MSX Band A (8.28 $\mu$m)
       (contours), 6 cm continuum peaks (triangles) 
       \citep{1986ApJ...307..302C}, 1.3 mm continuum emission clumps FIR2 to
       FIR6 (squares) \citep{1988A&A...191...44M}, and embedded infrared point
       sources IRS2 \citep{1974ApJ...193..373G} and IRS3 
       \citep{1989ApJ...342..883B} (stars).
       {\bf Bottom:} Overlay of the \iCII\ (color coded) and the 
       \mCII\ integrated intensities (contours: 100 K km/s to 625 K km/s in
       steps of 75 K km/s). The \iCII\ map
       has been smoothed to a resolution of 25$\arcsec$. The dashed box
       outlines the area 
       of strong emission that was used in the analysis of section
       \ref{discussion_13CII-fit}.
     }
\end{figure}

Figure \ref{int_intens_map} shows the measured \CII\ intensity distribution
integrated over the velocity interval 5--15 km/s. The emission agrees
remarkably well with the 8 $\mu$m MSX data
\citep{2001AJ....121.2819P,2008AJ....136.1947W}, which traces the UV-heated
dust in the interface between the \HII\ region and the molecular cloud. Only
the embedded MSX source just north of the map center is not very prominent in
the \CII\ data. The emission peaks along the southern rim of the \HII\ region,
where the interface is seen edge-on in a sharp ionization front.

The intensity distribution does not reflect the distribution of the molecular
material located in a roughly north-south elongated molecular ridge -- outlined
by the FIR peaks in Fig.~\ref{int_intens_map}. 

Averaged over a 60\arcsec\ diameter, we obtain a peak surface brightness of
4.4$\times$10$^{-3}$ ergs cm$^{-2}$ s$^{-1}$ sr$^{-1}$, in reasonable agreement
with KAO and ISO data \citep{1994ApJ...436..203J,2000A&A...358..310G}. 

\subsection{{\rm \iCII} distribution}
With a velocity span of $\pm$100 km/s, the spectra also cover the three
hyperfine structure components of the \iCII\ transition ($v_{\rm LSR}$-offsets
$-$65.2, 11.2, 63.2 km/s \citep{1986ApJ...305L..89C}) (Fig.~\ref{spectrum}). If
we co-add the velocity ranges [$-$60,$-$50], [19,25], and [70,80] km/s (taking
into account the systemic velocity), we find emission with a similar spatial
distribution as the main isotope (Fig.~\ref{int_intens_map}).

The \iCII\ emission is only prominent in a crescent-shaped area following the
edge-on ionization front, where column densities are highest. The peak emission
coincides with the \mCII\ emission maximum, but the \iCII\ intensity falls off
more rapidly, both to the \HII\ region side (north) and to the molecular cloud
side (south). The FWHM of the \iCII\ emission zone is $\approx$1\arcmin,
corresponding to a linear size of $\approx$0.1 pc.

\begin{figure}[bp]
    \begin{center}
      \includegraphics[width=6.5cm,angle=-90]{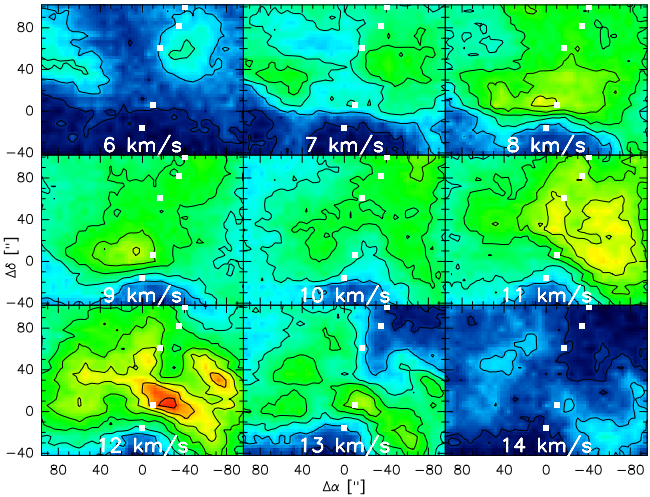}
     \end{center}
     \caption{
       \label{channel_maps}
       \mCII\ channel maps of 1 km/s wide velocity channels spaced by 1 km/s. 
       Panels are labeled by their center velocity. The markers denote the 
       positions of the clumps FIR2 to FIR6 \citep{1988A&A...191...44M}.
       Contour levels are from 10 K km/s to 150 K km/s in steps 
       of 20 K km/s.
     }
\end{figure}

\subsection{{\rm \mCII} channel maps}
\begin{figure}[bp]
    \begin{center}
      \includegraphics[width=6cm,angle=-90]{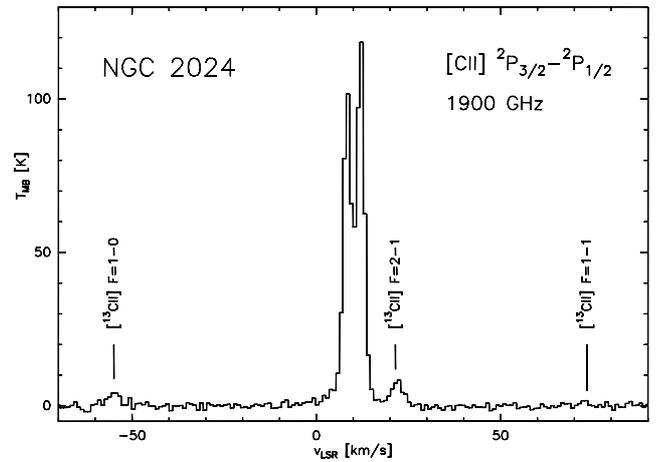}
     \end{center}
     \caption{
       \label{spectrum}
       Spectrum averaged over the area outlined in Fig.~\ref{int_intens_map}
       with marks indicating the predicted velocities of the 
       \iCII\ hyperfine components \citep{1986ApJ...305L..89C}.
    }
\end{figure}

The \CII\ emission is separated into two distinct velocity components, which
show up as two peaks in most spectra (Fig.~\ref{spectrum}).
Fig.~\ref{channel_maps} presents channel maps of 1 km/s wide velocity bins
ranging from 6 km/s to 14 km/s. The maps clearly show the two main velocity
peaks, which are predominant throughout most of the source. The low velocity
peak dominates the 8--9 km/s panels, the high velocity peak being clearly
visible at 11 and 12 km/s. The local minimum in the 10 km/s velocity bin around
FIR5 coincides with the peak of the integrated intensity map
(Fig.~\ref{int_intens_map}).

There appears to be an anticorrelation between the low velocity peak and the
high velocity peak. The relatively high intensities found at 9 km/s west of
FIR2 to FIR4 and east of FIR5 correspond to areas of weak emission at 12--13
km/s.

\section{Discussion}
\label{discussion_13CII-fit} 

For the data analysis, the interpretation of the two velocity peaks of the
spectra is crucial. Although the \CII\ peaks -- at velocities of $\sim$8.4 km/s
and $\sim$12 km/s -- are slightly offset from the two emission components in
the optically thin lines of very rare CO isotopes (9.2 km/s and 11.2 km/s,
\cite{1993ApJ...405..249G}), it seems reasonable to assume that the \CII\ line
may also be caused by two emission components and not by self-absorption. The
\iCII\ measurement allows us to clearly distinguish between these two
possibilities.

\subsection{{\rm \iCII} spectrum}

\begin{figure}[thbp]
    \begin{center}
      \includegraphics[width=6cm,angle=-90]{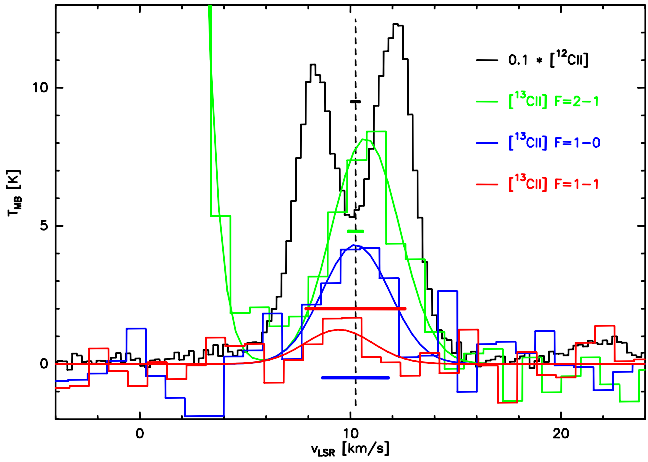}
     \end{center}
     \caption{
       \label{HFS_spectra}
       Spectrum of Fig.~\ref{spectrum} replotted on a common $v_{\rm LSR}$ 
       scale and overlaid by the model fit (see text and
       Table \ref{HFS_fit-results}). The dashed line marks the center velocity
       of the \mCII\ background component. Horizontal bars indicate
       the uncertainty in the rest frequencies. The \mCII\ 
       has been divided by ten.
    }
\end{figure}

To obtain a high signal-to-noise spectrum to analyze the isotopic line, we
averaged the spectra within a rectangular area encompassing the peak of the
\iCII\ emission (marked by a dashed box in Fig.~\ref{int_intens_map}).  The
resulting spectrum is shown in Fig.~\ref{spectrum}.  In Fig.~\ref{HFS_spectra},
we compare the line profile of the \mCII\ line to the three hyperfine
satellites of \iCII.

It is immediately obvious that the isotopic line only shows a single peak.
Thus, either a) only one of the two velocity components seen in \mCII\ has a
sufficiently high column density to produce measurable \iCII\ emission, or b)
the shape of the \mCII\ is produced by a warm broad background component
shadowed by a narrow cooler foreground component. Explanation a) would imply
that there is an error in the rest frequencies of $\ga$10 MHz, clearly beyond
the confidence limits of \citet{1986ApJ...305L..89C} and in contrast to the
good agreement found in recent HIFI measurements toward the Orion bar
\citep{2011IAUS..280P.280O}.  Thus, if we accept the rest frequencies to their
published accuracy, {\bf we are forced to the conclusion that the \mCII\ line
profile is caused by self-absorbed strong background emission} centered at
$v_{\rm LSR} \approx 10$~km/s.

Self-absorption in the \mCII\ line offers a quite natural explanation of the
anticorrelation observed in the spatial distribution of the two velocity peaks:
A velocity shift of the absorbing component with respect to the background
component shifts the observed line intensity from one velocity peak to the
other.

That the highest contrast between the 10 km/s dip and the 8 km/s and 12 km/s
peaks is found in an area of known strong foreground absorption
\citep{1986ApJ...307..302C,1989ApJ...342..883B} also strengthens this
interpretation. Radio recombination lines of carbon \citep[and reference
therein]{1982A&AS...48..345K} are also mostly near 10 km/s.

The discrepancy between the CO velocities and the self-absorbed \CII\ emission
indicates that the simple source model with only two velocity components may
underestimate the kinematic complexity of the source. 

\begin{table}[bp]
  \caption{
    \label{HFS_fit-results}
    Parameters of the fit to the spectrum of Fig.~\ref{HFS_spectra}.
    $\delta v_{\rm LSR}$ denotes the velocity offset of the
    \iCII\ hyperfine satellites to
    the \mCII\ background component. Bracket symbols: []: fixed value; \{\}: literature 
    value \citep{1986ApJ...305L..89C}; (): weakly constrained value.
  }
  \begin{center}
  \begin{tabular}[t]{cccccc}
     \hline
     \hline
     & \multicolumn{2}{c}{\mCII} & \multicolumn{3}{c}{\iCII} \\
     & emiss. & absorb. & $1\to0$ & $2\to1$ & $1\to1$ \\
     \hline
     $T_{\rm MB}$ [K]           & [400] & $-$352.5 &   4.3   &   8.2  &   (1.2) \\
     $v_{\rm LSR}$ [km/s]       & 10.26 & 10.20 &  $-$54.9  &  21.9  &  (72.7) \\
     $\Delta v_{\rm LSR}$ [km/s] & 3.8 & 2.7 &  [3.8]  &  [3.8]  &  [3.8] \\
     $\delta v_{\rm LSR}$ [km/s] & --- & --- &  $-$65.2  &  11.7  &  (62.4) \\
                       & --- & --- &  \{$-$65.2\}  &  \{11.2\}  &  \{63.2\} \\
     rel. intens.         & --- & --- &   0.31   &   0.60  &   (0.09) \\
               & --- & --- &   \{0.356\}   &   \{0.444\}  &   \{0.200\} \\
     \hline
  \end{tabular}
  \end{center}
\end{table}

\subsection{Hyperfine line ratios}
Figure \ref{HFS_spectra} includes the result of a five-component Gaussian fit
to the spectrum. The first two components model the main isotope (fit not shown
in Fig.~\ref{HFS_spectra}) as an emissive and an absorptive component.  The
remaining three components fit the three hyperfine lines, where the line width
is kept fixed and equal to the width of the broad background component. Table
\ref{HFS_fit-results} lists the fit results and compares the parameters of the
\iCII\ hyperfine satellites to the literature values.

For the two stronger hyperfine components, the velocities are within 0.5 km/s
(3 MHz) of the predicted values. The signal-to-noise ratio of the $F=1\to1$
component is too low to make a firm statement. The predicted line strength
ratio, however, is not reproduced by our measurement.  The
($F=2\to1$)/($F=1\to0$) intensity ratio should be 1.25, but is found to be 1.9.
A similar discrepancy was also found by \citet{2011IAUS..280P.280O}.

Obviously it cannot be ruled out that an as yet unknown emission line may
coincide with the $F=2\to1$ line, leading to excess emission at this velocity.
However, strong spectral lines at these wavelengths are so rare that this seems
highly unlikely. At the moment, we can only speculate that the anomalous
hyperfine ratio may be due to some non-LTE excitation mechanism, possibly
caused by hyperfine-selective radiative pumping by emission lines of the main
isotope at shorter wavelengths.

\subsection{Physical conditions}
The integrated intensity in the three hyperfine components adds up to 52 K
km/s, which -- in the high density limit ($n > 3000$~cm$^{-3}$) -- converts
into a column density $N$($^{13}$C$^+$) = $2.6\times10^{17}$ cm$^{-2}$
\citep{1985ApJ...291..755C}. Assuming standard abundance
ratios\footnote{Assumed abundances: $^{12}$C/$^{13}$C = 60; $^{12}$C/H =
10$^{-4}$}, this corresponds to $N$($^{12}$C$^+$) = $1.6 \times 10^{19}$
cm$^{-2}$ and a total hydrogen column density of $N$(H) = $1.6 \times 10^{23}$
cm$^{-2}$ in the gas emitting the \CII\ lines, if all the carbon is ionized.

Owing to the foreground absorption, the intrinsic \mCII\ emission is not
constrained well. The background emission line component requires temperatures
of several 100 K, with 165 K being a hard lower limit\footnote{Owing to the
Rayleigh-Jeans correction, the peak $T_{\rm MB}$ of 125 K
(Fig.~\ref{HFS_spectra}) requires gas temperatures of $\ge$165 K}. For a gas
temperature of e.g. 400 K, the \mCII\ line would be almost optically thick
($\tau\approx 2.4$) with an intrinsic line temperature of $T_{\rm
MB}$(\mCII,background) $\approx$ 325 K.

Standard spherical photon-dominated region (PDR) models
\citep{1996A&A...310..592S,2006A&A...451..917R} of even the most extreme single
clumps fall short by about one order of magnitude of reproducing this high line
intensity and large column density of ionized carbon. It is plausible that the
edge-on geometry of the ionization front may significantly boost the total
column density. Thus, an equivalent of ten extreme clumps would be required to
model the observed emission.  Similarly, in a clumpy PDR model
\citep{2008A&A...488..623C}, where the amount of material exposed to UV
radiation is drastically higher, the predicted \CII\ emission is much stronger.
For instance, a typical model\footnote{The single line data are insufficient to
fully constrain a model} with a total mass within the beam of 1 M$_{\odot}$, a
radiation field of $\chi = 10^3$ and an average density of 10$^6$ cm$^{-3}$ can
produce the observed \iCII\ intensity. These values correspond to a hydrogen
column density of $\approx$10$^{23}$ cm$^{-2}$, in agreement with the above
estimate.

With the absorption dip reaching down to 53 K (Fig.~\ref{HFS_spectra}), the
temperature in the foreground gas is limited to $T_{\rm FG} \le 90$
K\footnote{Rayleigh-Jeans correction: $T_{\rm{R-J}}$(90 K, 1900 GHz) = 53 K.}.
The optical depth derived for the foreground gas is $\tau_{\rm FG} \ge 1$,
where the exact value depends on the assumptions made about the temperatures.
If we accept an intrinsic main beam brightness temperature of 325 K for the
background component, the column density of ionized carbon in the absorbing
layer needs to be $N$(C$^+$) $\ge10^{18}$ cm$^{-2}$ for any gas temperature
above 40 K.

\section{Conclusions}
We have observed both \mCII\ and \iCII\ emission toward NGC 2024. The spatial
distribution of the emission follows the 8 $\mu$m continuum emission, tracing
the interface between the \HII\ region and the molecular cloud. The isotopic
line emission detected from an area near the ionization front implies that: 
\begin{itemize}
  \item{The double-peaked spectrum of the \mCII\ line is due to strong
background emission absorbed by cooler foreground gas at the same velocity.}
  \item{The kinematic signature of the ionized gas differs from that of the
molecular gas as traced by rare isotopes of CO.} 
  \item{The total hydrogen column density of warm gas at several hundred Kelvin 
traced by the \CII\ emission amounts to $\ge$10$^{23}$ cm$^{-2}$, suggesting
that the interface is highly clumpy.} 
  \item{The strong foreground absorption implies a hydrogen column 
density $\ge$10$^{22}$ cm$^{-2}$ of gas at temperatures below 100~K.}
\item{The measured intensity ratio of the hyperfine components does not match
the theoretical value, suggesting that there is some contribution from a
non-LTE excitation effect}
\end{itemize}

%----- acknowledgement
\begin{acknowledgements}
We thank Volker Ossenkopf and Markus R\"ollig for useful discussions.

We also thank the SOFIA engineering and operations teams whose tireless
support and good-spirit teamwork has been essential for the GREAT
accomplishments during Early Science, and say Herzlichen Dank to the DSI
telescope engineering team.

This work is based on observations made with the NASA/DLR Stratospheric
Observatory for Infrared Astronomy. SOFIA Science Mission Operations are
conducted jointly by the Universities Space Research Association, Inc., under
NASA contract NAS2-97001, and the Deutsches SOFIA Institut under DLR contract
50 OK 0901.  

\end{acknowledgements}
%----- bibliography

\bibliographystyle{aa}
\bibliography{AA_2012_18930}
\end{document}